\documentclass[prb,aps,twocolumn,a4paper,showpacs,citeautoscript,floatfix]{revtex4}

\usepackage{amsmath,amssymb,amsfonts}
\usepackage{bm}\let\vec\bm
\usepackage{graphicx}
\usepackage{color}

\begin{document}
\def\RH{R_{\text{H}}}

\title{Hall effect on the triangular lattice}

\author{G. Le\'{o}n}
\author{C. Berthod}
\author{T. Giamarchi}
\affiliation{DPMC-MaNEP, University of Geneva, 24 quai
Ernest-Ansermet, 1211 Geneva 4, Switzerland}
\author{A. J. Millis}
\affiliation{Columbia University, 538 West 120th Street, New York, NY 10027, USA}

\begin{abstract}
We investigate the high frequency Hall effect on a two-dimensional triangular lattice with nearest-neighbor hopping and a local Hubbard interaction. The complete temperature and doping dependencies of the high-frequency Hall coefficient $\RH$ are evaluated analytically and numerically for small, intermediate, and strong interactions using various approximation schemes. We find that $\RH$ follows the semiclassical $1/qn^*$ law near $T=0$, but exhibits a striking $T$-linear behavior with an interaction- and doping-dependent slope at high temperature. We compare our results with previous theories as well as Hall measurements performed in the cobaltates.
\end{abstract}

\date{\today}
\maketitle

\section{Introduction}\label{sec:one}

The interpretation of the Hall resistivity is made difficult even in relatively simple metals by the connection between the Hall constant and the relaxation time on the Fermi surface \cite{ashcroft}.
In many cases the simple semiclassical expression $\RH=1/qn^*$ does not work and we must therefore deal with a transport measurement encoding much more information than just the effective density $n^*$ and the sign $q$ of the charge carriers.

In strongly correlated systems the Hall effect is even more difficult to interpret because interactions can have a large influence on the Hall resistivity. This influence moreover increases as the dimensionality of the systems decreases.
There have been various attempts to describe the Hall effect in strongly correlated models with different geometries in 2D \cite{Shastry_Hall,lange_hall_constant}, quasi-2D \cite{yakovenko_phenomenological_model_exp} and quasi-1D \cite{proceedings_ISCOM06_Hall,leon_hall} systems, but a general theoretical understanding is still lacking.

Among the strongly correlated systems the triangular lattice exhibits a unique property: it has the smallest possible closed loop with an odd number of steps (namely $3$). Anderson proposed that the model could have a spin-liquid ground state at commensurate fillings such as one electron per site \cite{anderson_hgtc_hubbard}. These peculiarities make the triangular lattice a very interesting system, which has been investigated extensively in the last decades. In particular, important differences between the Hall effect in the square and triangular lattices were pointed out in Ref.~~\onlinecite{Shastry_Hall}. Additionally, many studies have been motivated by the recent discovery of superconductivity in CoO$_2$ layered compounds (cobaltates), which are good realizations of an isotropic 2D triangular lattice \cite{Singh_band}, and in organic conductors of the BEDT family \cite{Seo_chemrev} where one finds various structures resembling the anisotropic triangular system.

Several Hall measurements have been undertaken in both organic superconductors \cite{Sushko_Hall_organics, Katayama_Hall_organics} and cobaltates, especially in the Na$_{x}$CoO$_2$ compound. In the latter, the anomalous linear increase of the dc Hall coefficient \cite{Choi_Infrared} and a recent infrared Hall measurement \cite{Choi_Infrared} have motivated further theoretical work on this issue \cite{Shastry_triangular_TJ_model, Haerter_Peterson_Shastry, Shastry_Curie_Weiss, Koshibae_Hall} to investigate the effect of correlations in this system and their contribution to the Hall coefficient, but many questions remain open. In the case of the organic conductors, the anisotropy present in the BEDT family complicates the problem further.

In the present work, we study theoretically the Hall effect in a 2D triangular lattice where electrons interact via an onsite Coulomb repulsion $U$. We calculate $\RH$ in the high frequency limit \cite{Shastry_Hall, lange_hall_constant} where the probing frequency $\omega$ is the largest energy scale of the problem, and we cover the whole range of interaction values using several approximation schemes.

The paper is organized as follows: in Sec.~\ref{sec:model} we introduce the model and the formalism used to compute the Hall coefficient $\RH$ at high frequency on the triangular lattice. In Sec.~\ref{sec:results} we present our analytical and numerical results obtained in the whole $(n,U,T)$ parameter space. We discuss in detail the doping and temperature dependence of $\RH$. Sec.~\ref{sec:discussion} is devoted to a discussion of our results and a comparison with other theoretical approaches as well as experimental measurements. Finally our conclusions are given in Sec.~\ref{sec:conclusion}, and the subsequent appendices collect technical details.

\section{Model and method} \label{sec:model}

\begin{figure}[tb]
\includegraphics[width=7.5cm]{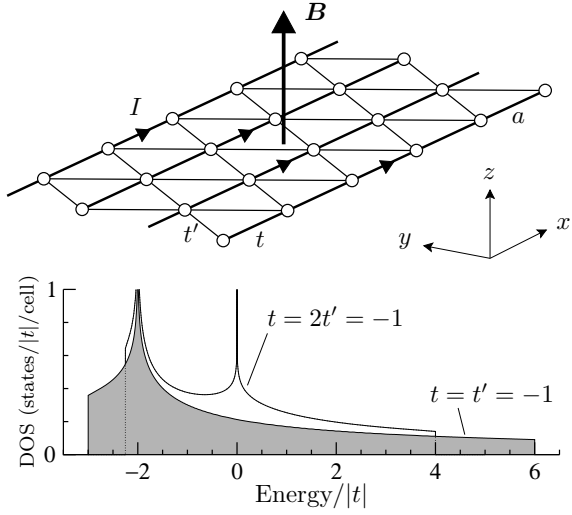}
\caption{\label{fig:model}
Top: Two-dimensional triangular lattice. $a$ is the lattice parameter, $t$ and $t'$ are the hopping amplitudes for bonds along the $x$ direction and for $\pm 60^\circ$ bonds, respectively.
The current $I$ flows along the $x$ axis,
the magnetic field $\vec{B}$ is applied along the $z$ axis, and the Hall voltage is
measured along the $y$ axis.
Bottom: Non-interacting density of states of the model in the cases $t=t'=-1$ and $t=2t'=1$. The DOS generically presents two van Hove singularities, which are degenerate in the isotropic lattice. The energy position of the van Hove singularity for $t=t'=-1$ corresponds to a band filling of $1/2$ electron per site ($1/4$ filling). }
\end{figure}

Our model is sketched in Fig.~\ref{fig:model}.
We consider an anisotropic triangular lattice with nearest-neighbor hopping amplitudes $t$ and $t'$ and an on-site Hubbard interaction $U$. The Hamiltonian reads:
\begin{equation}\label{H}
\mathcal{H}=- \sum_{\langle ij\rangle\sigma}t_{ij}c^{\dagger}_{i\sigma}
c^{\phantom{\dagger}}_{j\sigma} + U\sum_i n_{i\uparrow}n_{i\downarrow}
\end{equation}
where $c^{\dagger}_{\alpha}\,(c_\alpha)$ is the creation (annihilation) fermion operator, $n_\alpha$ is the fermionic number operator and $\langle ij\rangle$ are nearest-neighboring sites.
The dispersion relation for this model (Fourier transform of $t_{ij}$) is
\begin{equation}\label{dispertion}
\varepsilon_{\vec{k}} = -2t\cos(k_xa) -4t'\cos(k_xa/2)\cos(k_ya\sqrt{3}/2).
\end{equation}
The correponding density of states (DOS) exhibits two van Hove singularities which are degenerate when $t=t'$ (see Fig.~\ref{fig:model}). Unlike in the square lattice, the DOS has no particule-hole symmetry, irrespective of the value of $t$ and $t'$.

We assume that a current $I$ flows along the $x$ axis and a dc magnetic field $\vec{B}$ is applied along $z$, hence a Hall voltage develops along the $y$ axis (see Fig.~\ref{fig:model}). In order to represent the magnetic field and the applied ac electric field along $x$, we use the vector potential $\vec{A}=\vec{A}^{\text{mag}}+\vec{A}^{\text{el}}$, where for the magnetic part we choose the Landau gauge, $\vec{A}^{\text{mag}}=Bx\hat{\vec{y}}$, and $\vec{A}^{\text{el}}$ describes the electric field. The coupling between the lattice fermions and the electromagnetic field induces a Peierls phase in the hopping amplitudes which change according to $t_{ij}\to t_{ij}\exp(-ie\int_{i}^{j}\vec{A}\cdot\vec{dl})$.

The operator for the total current, $J_\mu=\int d\vec{r} j_\mu(\vec{r})$, and the diamagnetic susceptibilities $\chi_\mu(0)$ of the system are defined as usual:
\begin{eqnarray}
J_\mu&=&-S\sum_i\left.\frac{\delta\mathcal{H}}{\delta A_\mu(i)}\right|_{\vec{A}^{\text{el}}=0} \\ \chi_\mu(0)&=&-\frac{S}{N}\sum_i\left.\left\langle
\frac{\delta^2\mathcal{H}}{\delta A_\mu^2(i)}\right\rangle\right|_{\vec{A}=0} \nonumber \\ \label{diamag}
&=&-\frac{2e^2}{NS}\sum_{\vec{k}}\frac{\partial^2\varepsilon_{\vec{k}}}{\partial k_\mu^2}\langle n_{\vec{k}}\rangle
\end{eqnarray}
where $S=a^2\sqrt{3}/2$ is the unit-cell area, $NS$ is the total system surface, $\langle n_{\vec{k}}\rangle=\langle c^{\dagger}_{\bm{k}} c_{\bm{k}}\rangle$ is the distribution function and the thermodynamic average $\langle\cdots\rangle$ is taken with respect to the Hamiltonian of Eq.~(\ref{H}).

Performing the functional derivatives we find for the components of the currents
\begin{subequations}\label{currents}\begin{eqnarray}\nonumber
J_x&=&ea\Big[2t\sum_{\vec{k}\sigma}c^{\dagger}_{\vec{k}\sigma}c^{\phantom{\dagger}}_{\vec{k}\sigma}
\sin(k_xa)\\ \nonumber
&&+t'\sum_{\vec{k}\sigma}\sin\left(\frac{k_xa}{2}+\frac{\eta a}{4}\right)
\left(c^{\dagger}_{\vec{k}\sigma}c^{\phantom{\dagger}}_{\vec{k}+\vec{\eta}\sigma}
e^{ik_y\sqrt{3}\frac{a}{2}}+\text{h.c.}\right)\Big]\\
\\\nonumber
J_y&=&-ea\sqrt{3}t'\sum_{\vec{k}\sigma}\cos\left(\frac{k_xa}{2}+
\frac{\eta a}{4}\right)\\ \label{Jy}
&&\left(ic^{\dagger}_{\vec{k}\sigma}c^{\phantom{\dagger}}_{\vec{k}
+\vec{\eta}\sigma}e^{ik_y\sqrt{3}\frac{a}{2}}+\text{h.c.}\right)
\end{eqnarray}\end{subequations}
where we have defined the vector $\vec{\eta}=(\eta,0)$ with $\eta=\sqrt{3}eBa/2$.
The diamagnetic susceptibilities resulting from Eqs~(\ref{diamag}) and (\ref{dispertion}) are:
\begin{subequations}\begin{eqnarray}\label{Xx}\nonumber
\chi_x(0)&=&-\frac{4e^2}{\sqrt{3}}\frac{1}{N}\sum_{\vec{k}}\Big[2t\cos(k_xa)\\
&&+t'\cos\left(\frac{k_xa}{2}\right)
\cos\left(k_y\sqrt{3}\frac{a}{2}\right)\Big]\langle n_{\vec{k}}\rangle\\
\nonumber
\chi_y(0)&=&-\frac{4\sqrt{3}e^2t'}{N}\sum_{\vec{k}}\cos\left(\frac{k_xa}{2}\right)
\cos\left(k_y\sqrt{3}\frac{a}{2}\right)\langle n_{\vec{k}}\rangle.\\
\end{eqnarray}\end{subequations}

The Hall coefficient is defined as the ratio of the Hall resistivity to the applied magnetic field, $\RH=\rho_{yx}/B$, the Hall resistivity $\rho_{yx}$  being related to the conductivity tensor $\sigma_{\mu\nu}$ through
\begin{equation}
\rho_{yx} =\frac{\sigma_{xy}}{\sigma_{xx}\sigma_{yy}-\sigma_{xy}\sigma_{yx}}.
\end{equation}
As shown in Appendix \ref{app_memory}, it is possible to rewrite $\RH$ as a high-frequency series where the infinite-frequency limit reads
\begin{equation}\label{RH_highw}
\RH(\omega\to\infty)=\lim_{B\rightarrow 0}\left(-\frac{i}{B NS}\frac{\langle\left[J_x,J_y\right]\rangle}{\chi_x(0)\chi_y(0)}\right)
\end{equation}
and the remaining contributions are expressed in terms of a memory matrix \cite{gotze_fonction_memoire,lange_hall_constant}. Eq.~(\ref{RH_highw}) was originally derived in Ref.~~\onlinecite{Shastry_Hall} using a different frequency expansion. $\RH(\omega\to\infty)$ is expected to provide the dominant contribution at any finite frequency. The memory matrix formalism allows in principle to go beyond the infinite frequency approximation and compute corrections at finite frequency \cite{lange_hall_constant,leon_hall}. It leads, in particular, to corrections
due to interactions that vanish identically if $U=0$. These corrections do not affect the sign of $\RH$ which is entirely determined by $\RH(\omega\to\infty)$. In the following we shall consider only the infinite-frequency contribution to $\RH$, Eq.~(\ref{RH_highw}), and adopt the notation $\RH(\omega\to\infty)\equiv\RH$.

Strictly speaking, our results are valid provided the probing frequency is larger than any other energy scale in the system, $\omega > \max\lbrace U,t,T\rbrace$. The last two conditions, $\omega > \max\lbrace t,T\rbrace$, are easily fulfilled experimentally in known triangular compounds, while the condition $\omega>U$ is more problematic. However, as we will discuss in Sec. \ref{sec:discussion}, in certain limits our results coincide with those obtained in Ref.~~\onlinecite{Shastry_triangular_TJ_model} under the opposite assumption $\omega\ll U$, showing that this condition is not stringent.

In order to evaluate Eq.~(\ref{RH_highw}), we calculate the commutator $[J_x,J_y]$ from Eq.~(\ref{currents}), and we use the diamagnetic susceptibilities of Eq.~(\ref{diamag}) to arrive at
\begin{eqnarray}\label{RH}
\RH&=&\frac{S}{e}\frac{
\frac{1}{N}\sum_{\vec{k}}A_{\vec{k}}\langle n_{\vec{k}}\rangle}
{\frac{1}{N}\sum_{\vec{k}}B_{\vec{k}}\langle n_{\vec{k}}\rangle
\frac{1}{N}\sum_{\vec{k}}C_{\vec{k}}\langle n_{\vec{k}}\rangle},
\end{eqnarray}
with
\begin{eqnarray}\label{coeff}
\nonumber
A_{\vec{k}}&=&\cos\left(\frac{k_xa}{2}\right)\cos(k_xa)
\cos\left(k_y\sqrt{3}\frac{a}{2}\right)\\ \nonumber
&&+\frac{1}{4}(t'/t)\left[\cos(k_xa)+\cos\left(k_y\sqrt{3}a\right)\right]\\ \nonumber
B_{\vec{k}}&=&2\cos(k_xa)+(t'/t)\cos\left(\frac{k_xa}{2}\right)
\cos\left(k_y\sqrt{3}\frac{a}{2}\right)\\
C_{\vec{k}}&=&\cos\left(\frac{k_xa}{2}\right)
\cos\left(k_y\sqrt{3}\frac{a}{2}\right).
\end{eqnarray}
As can be seen from Eq.~(\ref{RH}) the high frequency Hall coefficient depends only on the distribution function $\langle n_{\vec{k}}\rangle$ as well as some geometrical factors. The interaction term in Eq.~(\ref{H}) therefore only influences $\RH$ through its effect on $\langle n_{\vec{k}}\rangle$. Since $\langle n_{\vec{k}}\rangle$ depends relatively weakly on $U$, we also expect the $U$-dependence of $\RH$ to be weak. Another implication of Eq.~(\ref{RH}) is that at low temparature the behavior of $\RH$ can be interpreted in terms of an effective carrier concentration, as in the non-interacting case.

\section{Results} \label{sec:results}

In the following we evaluate $\RH$ in the whole domain of interaction values $U$ with respect to the bandwidth $W=9t$ of the system, by using four differents approaches: exact calculation at $U=0$, a perturbative expansion of the self-energy at $U\lesssim W$, a local approximation to the self-energy, treated with dynamical mean field theory (DMFT) at $U\gtrsim W$, and finally the atomic limit of the self-energy at $U\gg W$.

\subsection{Non-interacting case}\label{sec:non-int}
In the non-interacting case there are various limits in which we can obtain analytical results for $\RH^0$ (\textit{i.e} $\RH$ at $U=0$): at zero temperature and band fillings near $n=0$ and $n=2$, and at high temperature $T\gg W$. For intermediate fillings and temperatures, we compute $\RH^0$ numerically by performing the sum in Eq.~(\ref{RH}) on a dense ($2048\times2048$) discrete $\vec{k}$-point mesh.

\subsubsection{Zero temperature}

Here we restrict for simplicity to the isotropic case $t'= t$ and we set the lattice parameter $a=1$.
Close to the band edges we can expand the various integrands of Eq.~(\ref{coeff}) and thus perform the $\vec{k}$ integrals.

Near the bottom of the band the Fermi surface is made of two nearly circular electron pockets around $(\frac{4\pi}{3},0)$ and $(\frac{2\pi}{3},\frac{2\pi}{\sqrt{3}})$. In each pocket we have $\xi_{\vec{k}}\equiv\varepsilon_{\vec{k}}-\mu\approx3t-\frac{3}{4}tk^2-\mu$, where $k$ is the momentum measured from the pocket center, and therefore $k_{\text{F}}^2=\frac{4}{3}(3-\mu/t)$. The corresponding electron density is $n=k_{\text{F}}^2/\pi$.
Writing similar expansions of $A_{\vec{k}}$, $B_{\vec{k}}$, and $C_{\vec{k}}$ close to the pocket center and performing the Brillouin zone integrations, we obtain the non-interacting Hall coefficient at low electron density:
\begin{equation}\label{RH0_T0_1}
\RH^0(T=0)=\frac{1}{ne}\left[1-\frac{3\pi n}{8}+\mathcal{O}(n^2)\right].
\end{equation}
At sufficiently low density we recover, in the above expression, the classical result $\RH^0=1/ne$.

Near the top of the band the Fermi surface is a nearly circular hole pocket centered at $\vec{k}=(0,0)$. Close to this point we have $\xi_{\vec{k}}\approx-6t+\frac{3}{2}tk^2-\mu$, and therefore $k_{\text{F}}^2=\frac{2}{3}(6+\mu/t)$. The corresponding density is obtained by substracting the contribution of the hole pocket from the maximum density: $n_h=2-k_{\text{F}}^2/2\pi$.
Similarly, for the functions $A_{\vec{k}}$, $B_{\vec{k}}$, and $C_{\vec{k}}$ we have to substract the contribution of the hole pocket from the contribution of the whole Brillouin zone, which turns out to be zero because
\begin{equation}\label{eq:zero}
\sum_{\vec{k}}A_{\vec{k}}=\sum_{\vec{k}}B_{\vec{k}}=\sum_{\vec{k}}C_{\vec{k}}=0.
\end{equation}
Thus, for low hole densities $n_h=2-n$ we find that the non-interacting Hall coefficient is given by
\begin{equation}\label{RH0_T0_2}
\RH^0(T=0)=-\frac{1}{n_he}\left[1-\left(\frac{\pi n_h}{4}\right)^2+\mathcal{O}(n_h^3)\right],
\end{equation}
and as $n_h\to0$ we have $\RH^0=-1/n_he$.

\begin{figure}[tb]
\includegraphics[width=6.5cm]{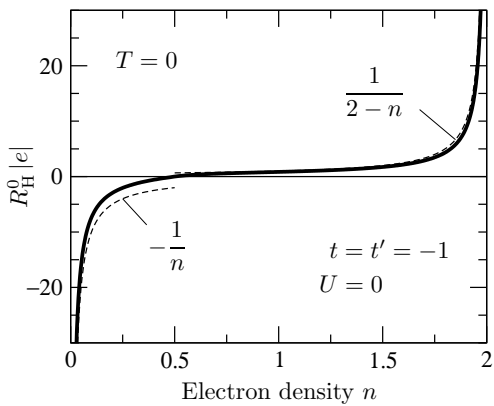}
\caption{\label{fig:RH0_T0}
Non-interacting Hall coefficient $\RH^0$ at zero temperature as a function of the electron density $n$, for an isotropic triangular lattice with $t'=t=-1$. The dashed lines indicates the clasiscal behavior at low electron and hole carrier densities.
}
\end{figure}

The complete density dependence of $\RH^0$ calculated numerically at zero temperature from Eq.~(\ref{RH}) is displayed in Fig~\ref{fig:RH0_T0} and compared to the limiting cases Eqs~(\ref{RH0_T0_1}) and (\ref{RH0_T0_2}). It is clear from this figure that the infinite-frequency $\RH$ follows the well-known dependence of the dc Hall coefficient $\RH(\omega=0)$ on the carrier charge density. This indicates a weak frequency dependence of the non-interacting Hall coefficient at zero temperature, since the dc result is recovered from the infinite frequency limit of $\RH$. Furthermore this suggests, as we will discuss in more details below, that the frequency dependence should not be too crucial, even in the presence of interactions, for most band fillings. At $U=0$ the sign of the Hall coefficient is entirely given by the sign of the carriers, and it can be seen from Fig.~\ref{fig:RH0_T0} how the sign changes when the Fermi energy crosses the van Hove singularity of the DOS, and the Fermi surface shape evolves from electron to hole like.

\subsubsection{High temperature}\label{section:RH0_highT}

\begin{figure}[tb]
\includegraphics[width=7cm]{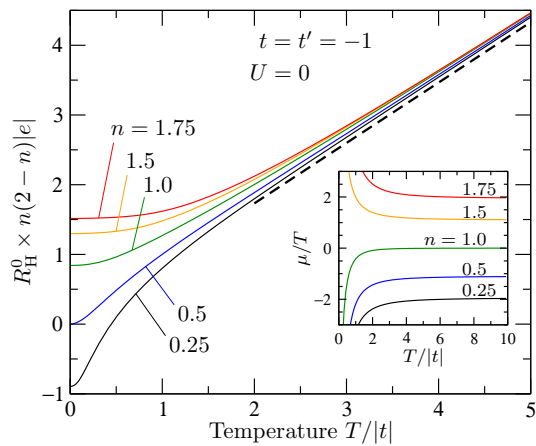}
\caption{\label{fig:RH_highT}
Temperature and electron density dependence of the non-interacting Hall coefficient $\RH^0$, for the isotropic triangular lattice with $t=t'=-1$. The dashed line shows the asympotic behavior described by Eq.~(\ref{RH-T}). Inset: Temperature and density dependence of the chemical potential $\mu$, illustrating the relation $\mu\sim T$ at high temperature.
}
\end{figure}

If $T\gg t$ the distribution function $\langle n_{\vec{k}}\rangle$, which reduces to the Fermi distribution at $U=0$, can be expanded in power of $\beta=1/T$. This expansion must be done at constant density $n$, which requires that $\beta\mu$ remains finite as $\beta\to 0$, in other words $\mu\sim T$ at high temperature. Taking this into account we can deduce the relation between $\mu$ and $n$, $\exp(-\beta\mu)=2/n-1$, and write the Fermi distribution as
\begin{equation}\label{expand}
\langle n_{\vec{k}}\rangle=\frac{n}{2}-n(2-n)\varepsilon_{\vec{k}}\frac{\beta}{4}
 + \mathcal{O}(\beta^2).
\end{equation}
Due to Eq.~(\ref{eq:zero}) the $\vec{k}$-independent terms in Eq.~(\ref{expand}) do not contribute to $\RH^0$, which thus takes the form:
\begin{equation}
\RH^0(T\gg t) = -4T\frac{S}{e}\frac{1}{n(2-n)}\frac{\frac{1}{N}\sum_{\vec{k}} A_{\vec{k}} \varepsilon_{\vec{k}}}{\frac{1}{N}\sum_{\vec{k}} B_{\vec{k}}\varepsilon_{\vec{k}} \frac{1}{N}\sum_{\vec{k}} C_{\vec{k}}\varepsilon_{\vec{k}}}.
\end{equation}
Performing the Brillouin zone integrations we obtain
\begin{equation}\label{RH-T}
\RH^0(T\gg t)=\frac{T/t}{e}\frac{1}{n(2-n)}\frac{a^2\sqrt{3}}{2}\frac{3}{2+(t'/t)^2}.
\end{equation}
This result is plotted in Fig.~\ref{fig:RH_highT} together with the numerically calculated full temperature and density dependence. The most striking feature of Eq.~(\ref{RH-T}) is the linear increase of $\RH^0$ with $T$. The same linear behavior was obtained in Ref.~~\onlinecite{Motrunich_Lee} at $\omega=0$, indicating a weak frequency dependence of $\RH^0$ at high temperature.
Our result shows that the $T$-linear dependence of $\RH$ is not due to interactions but to the peculiar topology of the triangular lattice. The sign of $\RH^0$ at high $T$ is determined by the sign of $t$, irrespective of the density (see Fig.~\ref{fig:RH_highT}). We attribute this property to the fact that at high enough temperatures the full band contributes to the Hall effect; hence the sign of $\RH$ reflects the dominant nature, electron or hole-like, of the band. As is clear from Fig.~\ref{fig:model}, for $t<0$ the band is dominantly hole-like, while for $t>0$ it is electron-like.

The relevance of result (\ref{RH-T}) is that even without interactions in the system, the Hall coefficient has a linear dependence at high temperature due to the geometry of the system, emphasizing the peculiarity of the triangular lattice.
By contrast on the square lattice the same analysis yields a $T$-independent non-interacting $\RH^0=\frac{2}{e}\left[\frac{1}{n}-\frac{1}{n(2-n)}\right]$ at high temperature.

\subsection{Weakly interacting regime}

\begin{figure}[b]
\includegraphics[width=8cm]{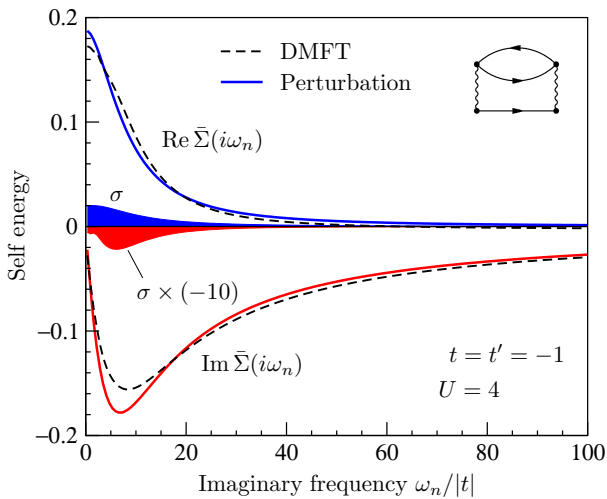}
\caption{\label{fig:Im_self}
Brillouin zone average of the real and imaginary parts of the self-energy Eq.~(\ref{self-E}) at low temperature $T=0.1$, calculated using a $64\times64$ $\vec{k}$-point mesh (solid lines). The small standard deviations $\sigma$ (shaded curves) illustrate the weak momentum dependence of $\Sigma(\vec{k},i\omega_n)$. The dashed lines show the local self-energy resulting from the DMFT calculation (see Sec.~\ref{section:DMFT}). The density was set to $n=1.54$, which is the value for Na$_{0.7}$CoO$_2$ (see Sec.~\ref{sec:discussion}). Inset: Feynman diagram corresponding to Eq.~(\ref{self-E}).}
\end{figure}

When interactions are present, the distribution function $\langle n_{\vec{k}}\rangle$ can be expressed in terms of the one-electron self-energy $\Sigma(\vec{k},i\omega_n)$ as:\cite{mahan_book}
\begin{equation}\label{n_k}
\langle n_{\vec{k}}\rangle=\frac{1}{\beta}\sum_{\omega_n}\frac{e^{i\omega_n0^+}}{i\omega_n-\xi_{\vec{k}}-\Sigma(\vec{k},i\omega_n)},
\end{equation}
with $\omega_n=(2n+1)\pi T$ the odd Matsubara frequencies. In the weak coupling regime $U\lesssim W$, we evaluate the self-energy using conventional perturbation theory in $U$ and we keep only the lowest order contributions of order $U^2$. For a local interaction like the Hubbard term in Eq.~(\ref{H}) there is only one diagram which is drawn in Fig.~\ref{fig:Im_self}. The standard diagrammatic rules yield the following expression for the self-energy:
\begin{eqnarray}\label{self-E}
&&\Sigma(\vec{k},i\omega_n)=-\frac{U^2}{N^2}\sum_{\vec{k}_1\vec{k}_2}\\\nonumber
&&\frac{f(\xi_{\vec{k}_2})\left[f(\xi_{\vec{k}_1})-f(\xi_{\vec{k}+\vec{k}_1-\vec{k}_2})\right]-f(\xi_{\vec{k}_1})
f(-\xi_{\vec{k}+\vec{k}_1-\vec{k}_2})}{i\omega_{n}+\xi_{\vec{k}_1}-\xi_{\vec{k}_2}-\xi_{\vec{k}+\vec{k}_1-\vec{k}_2}}
\end{eqnarray}
where $f(\xi_{\vec{k}})$ is the Fermi distribution function.

The numerical evaluation of Eq.~(\ref{self-E}) is demanding due to the double momentum integration. This is particularly time consuming because our calculations are done at fixed density, and thus require to calculate $\Sigma(\vec{k},i\omega_n)$ many times in order to determine the chemical potential. However it turns out that the momentum dependence of $\Sigma(\vec{k},i\omega_n)$ in Eq.~(\ref{self-E}) is weak.
This is illustrated in Fig.~\ref{fig:Im_self} where we plot the Brillouin zone average of the self-energy $\bar{\Sigma}(i\omega_n)$ as well as its standard deviation. The weak momentum dependence allows us to compute $\Sigma(\vec{k},i\omega_n)$ on a coarse (typically $16\times16$) $\vec{k}$-point mesh, and then to interpolate using splines onto a dense mesh for the evaluation of $\langle n_{\vec{k}}\rangle$ and eventually $\RH$. The Matsubara sum in Eq.~(\ref{n_k}) also requires special attention: the formal regularization of the divergence through the exponential factor is not suitable for a numerical evaluation of the sum. We therefore rewrite Eq.~(\ref{n_k}) as
\begin{equation}\label{n_k_2}
\langle n_{\vec{k}}\rangle=\frac{1}{2}+\frac{1}{\beta}\sum_{\omega_n}\left(\frac{1}{i\omega_n-\xi_{\vec{k}}-\Sigma(\vec{k},i\omega_n)}-\frac{1}{i\omega_n}\right).
\end{equation}
The $\omega_n$ sum is now convergent and can be efficiently calculated via the truncation at some large frequency and the analytical evaluation of the remaining terms using an asymptotic expansion of the self-energy.

The $\RH$ resulting from perturbation theory are valid in the regime $ U < W\ll\omega$, with $W=9|t|$ the bandwidth of the system.
As already anticipated the effect of a small $U$ on the distribution $\langle n_{\vec{k}}\rangle$ is a subtle broadening, and as a result the dependence of $\RH$ on $U$ is very weak at low $U$. Fig.~\ref{fig:RH_vs_U} provides an illustration of this weak dependence. As a consequence the non-interacting results of Sec.~\ref{sec:non-int} are expected to give a fairly good account of the Hall effect for an interaction strength smaller than the bandwidth $W$.

An important observation which we can make from our perturbative calculations is that the momentum dependence of the self-energy is very small, \textit{i.e.} the self-energy is almost local in real space. This suggests to approach the strong-coupling regime $U\gtrsim W$ by assuming that the self-energy is \emph{exactly} local. In the following section we study such local approximations to the self-energy, and we compare them to the result of the perturbation theory.

\subsection{Strongly interacting regime}
Assuming that the self-energy is local in first approximation, we investigate here two models for $\Sigma(i\omega_n)$ and their implications for the Hall coefficient $\RH$. The first approach is based on the dynamical mean field theory (DMFT) \cite{georges_dmft} and requires to solve a difficult self-consistent quantum impurity problem. Due to numerical difficulties this method cannot be pushed to very high interactions and/or very low temperature. Our second approach is based on a simple analytical form for $\Sigma(i\omega_n)$, which is expected to be valid at $U\gg W$, and allows us to express $\langle n_{\vec{k}}\rangle$ analytically in this limit.

\subsubsection{DMFT}\label{section:DMFT}

In the DMFT framework the local self-energy is expressed as:
\begin{equation}\label{sigma}
\Sigma(i\omega_n) = \mathcal{G}_0^{-1}(i\omega_n)-\mathcal{G}^{-1}(i\omega_n)
\end{equation}
where $\mathcal{G}_0$ is an effective propagator describing the time evolution of the fermions in the absence of interaction, and $\mathcal{G}$ is the full propagator, which takes into account the local Hubbard interaction. The calculation of $\mathcal{G}$ from a given $\mathcal{G}_0$ amounts to solve the problem of a quantum impurity embedded in a bath. We do it by means of the quantum Monte Carlo Hirsh-Fye algorithm \cite{hirsch_fye_qmc} as described in Ref.~~\onlinecite{georges_dmft}. From the requirement that $\mathcal{G}$ coincides with the local Green's function of the lattice, \textit{i.e.}
\begin{equation}
\mathcal{G}(i\omega_n)=\frac{1}{N}\sum_{\vec{k}}\frac{1}{i\omega_n-\xi_{\vec{k}}-\Sigma(i\omega_n)},
\end{equation}
one can deduce the self-consistency condition
\begin{equation}\label{eq:self-consistency}
\mathcal{G}_0^{-1}(i\omega_n)=1/\tilde{D}\left[i\omega_n-\Sigma(i\omega_n)\right]+\Sigma(i\omega_n),
\end{equation}
where $\tilde{D}(z)\equiv \int d\xi\,D(\xi)/(z-\xi)$ is the Hilbert transform of the DOS $D(\xi)$ corresponding to the triangular lattice and shown in Fig.~\ref{fig:model}. Once the self-consistent $\mathcal{G}_0(i\omega_n)$ is obtained, the corresponding self-energy $\Sigma(i\omega_n)$ is injected in Eq.~(\ref{n_k_2}) to compute $\RH$.

\begin{figure}[tb]
\includegraphics[width=7.5cm]{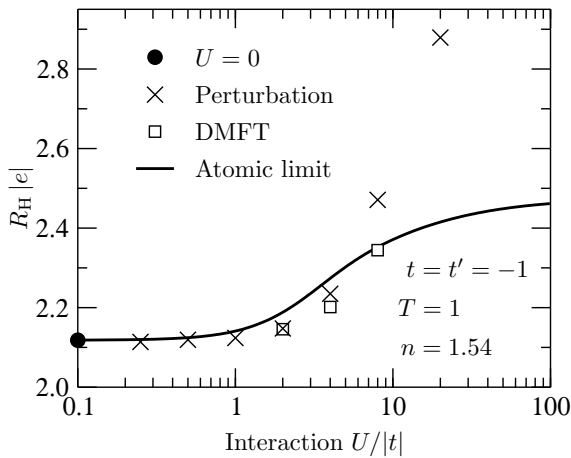}
\caption{\label{fig:RH_vs_U}
Evolution of the high-frequency Hall coefficient with $U$ calculated, for an isotropic triangular lattice $t=t'=-1$, using different approximations at $T=|t|$ and $n=1.54$}
\end{figure}

In Fig.~\ref{fig:Im_self} we compare the DMFT self-energy with the Brillouin zone average of the perturbative expression Eq.~(\ref{self-E}), both calculated at $U=4$. It can be seen that the frequency dependence and the order of magnitude of the two quantities are very similar, suggesting that the self-energy is dominated by the $U^2$ term and therefore the domain of validity of the perturbation theory is not limited to very small $U$. On the other hand it shows that the DMFT, although it is not a perturbative approach, provides a smooth transition from the weak to the strong-coupling regimes. This is further illustrated in Fig.~\ref{fig:RH_vs_U} where we see that the values of $\RH$ calculated by perturbation theory and DMFT coincide up to $U\approx4|t|$. At not too low temperature the DMFT calculation is reliable up to interaction strengths comparable to the bandwidth $W$. We have performed DMFT calculations at $U>W$, but since these results could be affected by systematic statistical errors in the Monte-Carlo summation, they are not shown in Fig.~\ref{fig:RH_vs_U}. At $U\gg W$ it is expected that the DMFT result approaches the atomic limit in which accurate calculations can be performed, as discussed in the next paragraph.

\subsubsection{Atomic limit}\label{atomic_limit}

In the limit of very strong interactions $U\gg W$ we assume that the self-energy approaches its atomic limit given by the expression (see Appendix \ref{app_atomic}):
\begin{equation}\label{atomic-self}
\Sigma_{\text{at}}(i\omega_n)=\frac{nU}{2}+\frac{n/2(1-n/2)U^2}{i\omega_n+\mu_{\text{at}}-(1-n/2)U}
\end{equation}
with $\mu_{\text{at}}$ the chemical potential in the atomic limit, not to be confused with the lattice chemical potential $\mu$.
Using this expression in Eq.~(\ref{n_k}) it is possible to evaluate analytically the sum over Matsubara frequencies and thus to obtain a closed expression for $\langle n_{\vec{k}}\rangle$ (Appendix \ref{app_atomic}).
In Fig.~\ref{fig:RH_vs_U} we show the Hall coefficient calculated with the atomic limit of the self-energy in the whole range of interaction values. $\RH$ obviously converges to the non-interacting limit at low $U$ since the atomic self-energy vanishes at $U=0$, and provides a good interpolation between the weak and the strong-coupling regimes. At intermediate values $U\sim W$ the atomic limit is not reliable, although it gives the correct order of magnitude for $\RH$. Fig.~\ref{fig:RH_vs_U} also shows that $\RH$ saturates at sufficiently large $U$.

\begin{figure}[b]
\includegraphics[width=7cm]{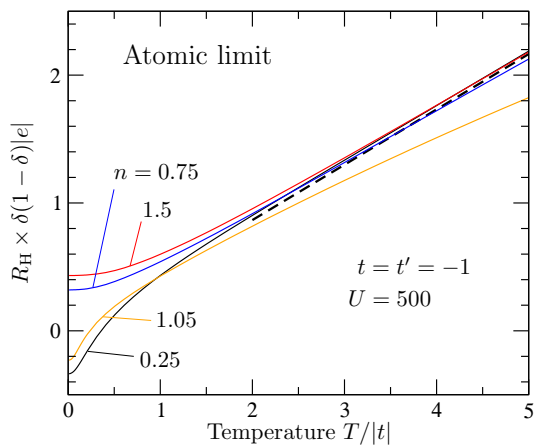}
\caption{\label{fig:atomic_limit}
Temperature and density dependence of the high frequency Hall coefficient calculated in the atomic limit at $U=500|t|$. The dashed line shows the asymptotic behavior, Eq.~(\ref{RHinf-T}), and $\delta =|n-1|$. For densities close to half-filling ($n=1.05$), $\RH(T)$ deviates from the asymptotic behavior (see text).}
\end{figure}

In Fig.~\ref{fig:atomic_limit} we display the temperature and density dependence of $\RH$ at $U=500|t|$, which is a value typical for the cobaltate compounds as discussed in the next section. We have selected four densities corresponding to the bottom and top of the lower and upper Hubbard bands (see also Fig.~\ref{fig:hubbard_bands} below). Like for $U=0$ we find a $T$-linear increase of $\RH$ at $T\gtrsim W$. Due to the Mott gap, however, the density dependence of the slope is not the same as for $U=0$. The slope can be obtained explicitly by sending $U$ to $+\infty$ and performing the high-temperature expansion as in Sec.~\ref{section:RH0_highT} (see Appendix~\ref{app_atomic}). The result is
\begin{equation}\label{RHinf-T}
		\RH^{U=\infty}(T\gg t)=\frac{T/t}{e}\frac{1}{\delta(1-\delta)}
		\frac{a^2\sqrt{3}}{4}\frac{3}{2+(t'/t)^2},
	\end{equation}
very similar to Eq.~(\ref{RH-T}) except that the slope $\propto[4\delta(1-\delta)]^{-1}$ replaces $[2n(2-n)]^{-1}$, where $\delta=|n-1|$ measures the departure from half-filling. The $U=\infty$ result of Eq.~(\ref{RHinf-T}) is displayed in Fig.~\ref{fig:atomic_limit}, and correctly describes our high-temperature results at $U=500|t|$.
The differences observed at $n=1.05$ in Fig.~\ref{fig:atomic_limit} reflect the fact that close to half-filling the slope of the high temperature $\RH$ depends strongly on the interaction and is not saturated even at $U=500|t|$ (see also Fig.~7). Away from half-filling the $U$ dependence of the slope is weaker, and Eq.~(24) is valid for lower interaction values.

\section{Discussion}\label{sec:discussion}
The various approximations presented above allow us to calculate the Hall coefficient on the triangular lattice for all interactions strengths $U$ and all temperatures $T$. The main limitation of our approach, in view of a comparison with experimental systems, is that our results are in principle valid in the limit $W,U \ll \omega$, because they are based on a high-frequency expansion. The first criterion, $W \ll \omega$, is not too difficult to satisfy for realistic compounds if the measurement of the Hall effect is performed at optical frequencies. The second criterion, $U\ll\omega$, seems more problematic since interaction strengths can be as large as several eV, at the upper edge of the mid-ultraviolet frequency domain. However, we have seen (Fig.~\ref{fig:RH0_T0}) that at $U=0$ and $T=0$ the Hall coefficient calculated at $\omega=\infty$ coincides with the $\omega=0$ dc value, and at $U=0$ and $T\gg t$, we obtained the $\omega=0$ results of Ref.~~\onlinecite{Motrunich_Lee}. All this suggests that the frequency dependence of $\RH$ is weak in the non-interacting case.

\begin{figure}[tb]
\includegraphics[width=6cm]{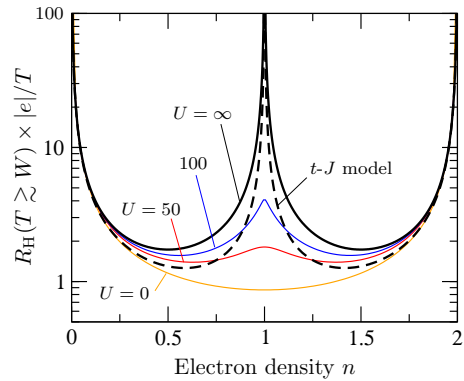}
\caption{\label{fig:comparison} Density dependence of the Hall coefficient within the $T$-linear regime at $U=0$ and in the atomic limit at $U\gg|t|$ (solid lines), compared to the result of the $t$-$J$ model (Ref.~~\onlinecite{Shastry_triangular_TJ_model}, dashed line).
}
\end{figure}

At the other extreme of the parameter space, $U=\infty$ and $T\gg W$, we can compare the result of the atomic limit approximation with the result of the $t$-$J$ model \cite{Shastry_triangular_TJ_model}. In the latter model $U$ is considered infinite from the outset, so that the high-frequency and high temperature expansion of Ref.~~\onlinecite{Shastry_triangular_TJ_model} is in fact valid at frequencies $\omega<U$. We plot in Fig.~\ref{fig:comparison} the density dependence of $\RH$ obtained in both models at $U=\infty$ and  $T\gtrsim W$. The small quantitative difference between the atomic limit at $U=\infty$ and the $t$-$J$ model shows that these two ways of treating the $U=\infty$ limit are not equivalent: they differ, in particular, in the renormalization of the kinetic energy by the interaction. However the two models give, where they can be compared, very similar behaviors. This reinforces the idea that the frequency dependence of $\RH$ is weak. Exact diagonalization on small clusters also indicate such a weak frequency dependence \cite{Haerter_Peterson_Shastry}. This strongly suggest that our results could also be valid at $\omega<U$, and therefore be relevant to interpret experiments performed in this regime.
The atomic-limit approach has the advantage to give access to the full temperature dependence (Fig.~\ref{fig:atomic_limit}) as well as the $U$-dependence as shown in Fig.~\ref{fig:comparison}, while the calculation of Ref.~~\onlinecite{Shastry_triangular_TJ_model} is valid at $U=\infty$ and $T\gg W$.

The evolution of $\RH$ with temperature is of particular interest since it is most easily probed experimentally. A linear increase of $\RH$ with temperature, without saturation at high $T$, was reported in Ref.~~\onlinecite{Shastry_triangular_TJ_model} for the $t$-$J$ model. Our results show that the Coulomb interaction is not responsible for this effect which is also present at $U=0$ (Fig.~\ref{fig:RH_highT}) and is therefore a consequence of the peculiar geometry of the triangular lattice. However the interaction controls the density dependence of the slope which changes smoothly from $[2n(2-n)]^{-1}$ at $U=0$ to $[4\delta(1-\delta)]^{-1}$ at $U=\infty$. This is further corroborated in Fig.~\ref{fig:comparison}.

\begin{figure}[b]
\includegraphics[width=6cm]{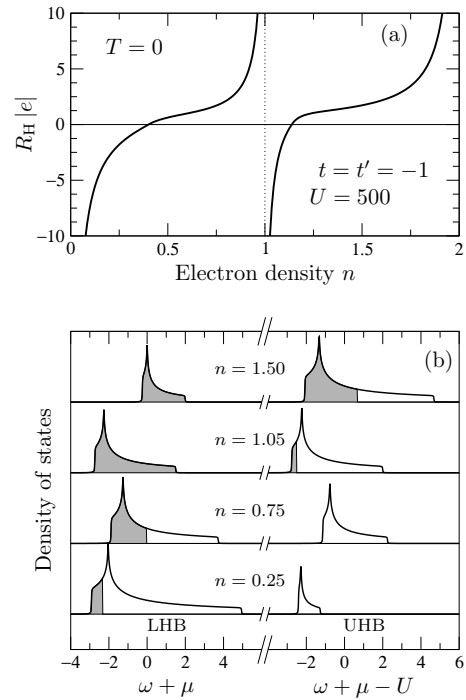}
\caption{\label{fig:hubbard_bands} (a) Density dependence of the Hall coefficient at $T=0$ and $U=500|t|$ calculated in the atomic limit approximation. (b) Density of states for various electron densities, showing the lower (LHB) and upper (UHB) Hubbard bands. The shaded regions indicate the occupied states and the position of the chemical potential.
}
\end{figure}

The sign of $\RH$ turns out to be independent of $n$ and $U$ at high temperature, unlike in the square lattice where $\RH$ changes sign at $n=1$. The situation is different at $T=0$. In the non-interacting case $\RH$ changes sign at quarter filling and can be simply interpreted in terms of the carrier density (Fig.~\ref{fig:RH0_T0}). We have also investigated the $T=0$ density dependence of $\RH$ at large $U$ as shown in Fig.~\ref{fig:hubbard_bands}. The interpretation in terms of the carrier density remains qualitatively valid, provided one takes into account the splitting of the DOS into the lower and upper Hubbard bands. These two bands are displayed in Fig.~\ref{fig:hubbard_bands}b, where it can also be seen that the DOS keeps qualitatively the same shape as for $U=0$, but the width of each band varies strongly with the density $n$. Due to this band renormalization the sign change of $\RH$ at $n<1$ does not occur at quarter filling, but a little below. Comparing Fig.~\ref{fig:hubbard_bands}a with Fig.~\ref{fig:comparison} one easily understands why the temperature dependence of $\RH$ has to be more pronounced slightly above $n=0$ and $n=1$ than slightly below $n=1$ and $n=2$, as can be also seen in Fig.~\ref{fig:atomic_limit}.

Among the Hall measurements reported for layered compounds with a triangular lattice structure, there is one performed at finite frequency by Choi \textit{et al.} \cite{Choi_Infrared} on the cobaltate Na$_{0.7}$CoO$_2$. This material is composed of two-dimensional layers of edge-sharing CoO$_6$ octahedra separated by an insulating layer of Na$^+$ ions, leading to a triangular lattice of CoO$_2$ units \cite{Son_NaxCoO2}. ARPES measurements \cite{Hasan_ARPES} indicate that the triangular lattice is isotropic with an estimated hopping amplitude of $t=-10$~meV and an effective Hubbard energy $U\sim5$ eV. From the radius of the Fermi-surface hole pocket observed in ARPES,
$k_{\text{F}}=0.65\pm0.1$~\AA$^{-1}$, we deduce an electron density $n=1.54$. Choi \textit{et al.} measured the temperature dependence of both the dc and ac Hall coefficients up to room temperature. The ac measurement was performed at $\omega=1100$~cm$^{-1}\approx12|t|$. The experimental conditions thus satisfy $T, W<\omega\ll U$.

\begin{figure}[tb]
\includegraphics[width=8cm]{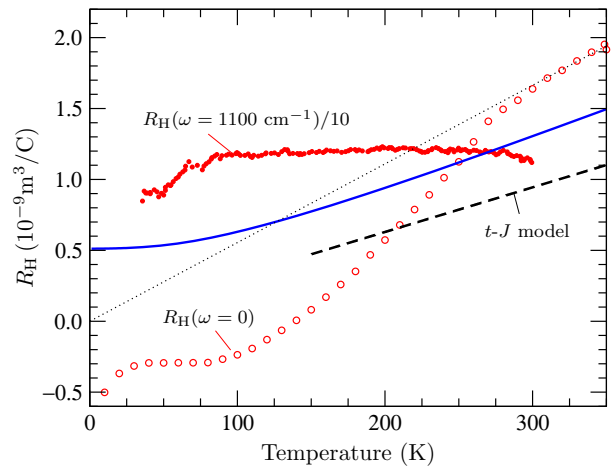}
\caption{\label{fig:cobaltates} Comparison of the Hall coefficient in Na$_{0.7}$CoO$_2$ as measured by Choi \textit{et al.} \cite{Choi_Infrared} at $\omega=0$ (empty circles) and $\omega=1100$~cm$^{-1}$ (full circles) with available theoretical models. The blue solid line is the high-frequency result in the atomic limit and the dashed line shows the high-temperature result in the $t$-$J$ model \cite{Shastry_triangular_TJ_model}. Theoretical curves are calculated at $t=t'=-10$~meV, $U=5$~eV, and $n=1.54$. The dotted line shows the best fit of the experimental data at high temperature with theoretical models (see text).}
\end{figure}

We note, however, that there are discrepancies between different sets of experimental data \cite{Choi_Infrared, Wang_Hall}.
The behavior of the dc $\RH$ above $T=250$~K is consistent with the linear increase predicted by the various theoretical models. By adjusting this models on the dc experimental data at high temperature (dotted line on Fig.~\ref{fig:cobaltates}) we obtain an independent determination of the hopping amplitude $t$, namely $t=-7.4$~meV using the atomic limit model Eq.~(\ref{RHinf-T}) and $t=-5.7$~meV using the $t$-$J$ model. This values are in good agreement with the ARPES determination of $t$.

The organic conductors of the BEDT family present several compounds with an anisotropic triangular structure. Unfortunately we are not aware of any measurements of the ac Hall effect which we could compare to our calculations, although measurements have been done at zero frequency in these materials.\cite{Sushko_Hall_organics, Katayama_Hall_organics}

\section{Conclusion} \label{sec:conclusion}

The theoretical high-frequency Hall coefficient in the two-dimensional triangular lattice exhibits two different characteristic behaviors at low and high temperatures: near $T=0$, $\RH$ resembles the classical dc Hall coefficient $1/qn^*$ where $q$ and $n^*$ are the carrier charge and density, respectively; at temperatures higher than the bandwidth, on the other hand, $\RH$ shows a remarkable $T$-linear behavior with a density- and interaction-independent slope. These conclusions apply provided the probing frequency is larger that the other energy scales of the problem, and that the electron self-energy remains essentially local for strong interactions.

Although we do not expect that a possible momentum dependence of the self-energy can have a strong effect on $\RH$, and we have argued that the frequency dependence of $\RH$ is probably weak, it is clear that for understanding the anomalously large $\RH(\omega)$ measured experimentally in Na$_{0.7}$CoO$_2$ in the mid-infrared range, one would have to extend the approach in order to cover the domain of intermediate frequencies. Another possibility is that the simple one-band model considered in this study would not suffice to capture the detailed properties of the materials \cite{Weber_triangular}. Experiments conducted as a function of $\omega$, as well as measurements of other materials with a triangular structure, would be very helpful to elucidate the peculiarities of the Hall effect in triangular compounds.

\acknowledgments

The authors are grateful to B. S. Shastry and H. D. Drew for valuable discussions. This work was supported by the Swiss National Science Foundation through Division II and MaNEP, and by Grant No. NSF-DMR-0707847.

\appendix

\section{Memory matrix formalism}
\label{app_memory}

The use of the memory matrix allows one to perform finite-order frequency expansions, which are singular for the conductivities due to their resonance structure \cite{gotze_fonction_memoire}. This approach has been used in previous works to study transport properties in Luttinger liquids \cite{giamarchi_umklapp_1d}, as well as the Hall effect in the 2D Hubbard model \cite{lange_hall_constant} and in quasi one-dimensional systems \cite{leon_hall}.

As we want to calculate the Hall resistivity $\rho_{yx}$, we start from the general relation between $\rho_{yx}$ and the conductivity tensor
$\sigma_{\mu\nu}$:
	\begin{equation}
		\rho_{yx}=\frac{\sigma_{xy}}{\sigma_{xx}\sigma_{yy}+\sigma_{xy}^2}.
	\end{equation}
Then we rewrite the conductivity tensor in terms of
a memory matrix $\bm{M}(\omega)$ as\cite{gotze_fonction_memoire}
\begin{equation}\label{sigma_M}
\bm{\sigma}^T(\omega)=i\left\{\omega\openone+\bm{\chi}(0)
\left[\bm{\Omega}+i\bm{M}(\omega)\right]\bm{\chi}^{-1}(0)\right\}^{-1}\bm{\chi}(0)
\end{equation}
where $\bm{\sigma}^T$ denotes the transpose of $\bm{\sigma}$, $\bm{\Omega}$ is called the frequency matrix, and $\bm{M}(\omega)$ is the memory matrix. $\bm{\chi}(0)$ is a diagonal matrix composed of the diamagnetic susceptibilities in each direction,
$\chi_{\mu\nu}(0)=\delta_{\mu\nu}\chi_{\mu}(0)$.
The frequency matrix $\bm{\Omega}$ in Eq.~(\ref{sigma_M}) is defined in terms of the equal-time current-current correlators as \cite{lange_hall_constant}
\begin{equation}\label{Omega}
\Omega_{\mu\nu}=\frac{1}{\mathcal{S}\chi_{\mu}(0)}
\left\langle\big[J_{\mu},J_{\nu}\big]\right\rangle.
\end{equation}
with $\mathcal{S}$ the sample surface (in a two-dimensional system).
Now we invert Eq.~(\ref{sigma_M}) and express the memory matrix $\bm{M}$ in terms of the conductivity tensor. For the Hall coefficient $\RH$ we need only the off-diagonal term $M_{xy}$ given by
\begin{equation}\label{M}
iM_{xy}(\omega)=\frac{i\chi_y(0)\sigma_{xy}(\omega)}
{\sigma_{xx}(\omega)\sigma_{yy}(\omega)+\sigma_{xy}^2(\omega)}-\Omega_{xy}.
\end{equation}
This implies that the Hall coefficient $R_{\text{H}}=\rho_{yx}/B$ can be expressed in term of the frequency and memory matrices as
\begin{equation}\label{RH_memory}
\RH(\omega)=\frac{1}{i\chi_y(0)}\lim_{B\to0}
\frac{\Omega_{xy}+iM_{xy}(\omega)}{B}.
\end{equation}
Since the memory matrix vanishes as $\omega^{-2}$ at high frequency, we see that the Hall coefficient is given by Eq.~(\ref{RH_highw}) in the infinite frequency limit.

\section{Calculation of the DMFT self-energy}\label{app_DMFT}

We evaluate the local self-energy in the DMFT framework using the Hirsh-Fye algorithm \cite{hirsch_fye_qmc} as described in Ref.~~\onlinecite{georges_dmft}. In this method the imaginary-time axis $[0,\beta[$ is cut into $L$ slices, and the Trotter formula is used in each time slice in order to single out the Hubbard interaction. In a second step the interaction is decoupled via the introduction of an Ising variable in every time slice. The Green's function $\mathcal{G}(\tau)$ is finally calculated by averaging over the ensemble of configurations of the Ising variables using a Monte-Carlo sampling and local updates. In our calculations at $\beta=1$ and $U\leqslant20$ we take $L=128$ and we keep $10^6$ out of the $\sim 10^8$ configurations visited. The numerical accuracy of the calculated $\mathcal{G}(\tau)$ is estimated to be $\sim10^{-3}$ at the highest $U$ values, and closer to $10^{-4}$ at $U\lesssim4$.

\begin{figure}[tb]
\includegraphics[width=8cm]{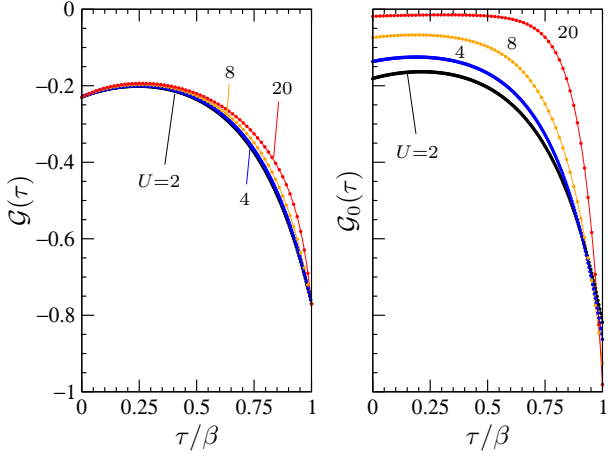}
\caption{\label{fig:DMFT}
DMFT imaginary-time propagators calculated at $\beta=1$ and $n=1.54$ for various interaction strengths. The symbols show the QMC results on the discrete-time mesh. The solid lines are the fit, Eq.~(\ref{eq:fit_poles}), used to evaluate the self-energy shown in Fig.~\ref{fig:Im_self}.
}
\end{figure}

In order to calculate the self-energy and solve the DMFT self-consistency condition, Eq.~(\ref{eq:self-consistency}), we need to Fourier transform $\mathcal{G}(\tau)$ from imaginary time to imaginary Matsubara frequencies $i\omega_n$. In traditional implementations of the algorithm this step is performed through a cubic spline interpolation of $\mathcal{G}(\tau)$. Because cubic splines are non-analytic, however, the resulting Fourier series are unreliable at frequencies above $\sim L/\beta$. Instead of an interpolation, we have performed a fit of $\mathcal{G}(\tau)$. Our fitting function is a discrete form of the spectral representation, $\mathcal{G}(\tau)=-\int d\varepsilon\,A(\varepsilon) e^{-\varepsilon\tau}f(-\varepsilon)$, which we express as
	\begin{equation}\label{eq:fit_poles}
		\mathcal{G}(\tau)=-\sum_{j=1}^{M}A_je^{-\varepsilon_j\tau}f(-\varepsilon_j)
	\end{equation}
with $A_j\geqslant0$ and $\sum_{j=1}^M A_j=1$. The number $M$ of poles $\varepsilon_j$ and their weight $A_j$ are determined by adding more and more terms in Eq.~(\ref{eq:fit_poles}), until the fitted function matches all QMC data points within a numerical tolerance, which we take as the estimated accuracy of $\mathcal{G}(\tau)$. The Fourier transform is then simply
	\begin{equation}\label{eq:continuation}
		\mathcal{G}(i\omega_n)=\sum_{j=1}^{M}\frac{A_j}{i\omega_n-\varepsilon_j}.
	\end{equation}
The calculated self-consistent propagators $\mathcal{G}(\tau)$ and $\mathcal{G}_0(\tau)$ are displayed in Fig.~\ref{fig:DMFT}, together with the fits to Eq.~(\ref{eq:fit_poles}).

Solving Eq.~(\ref{eq:self-consistency}) at fixed electron density requires to determine the chemical potential $\mu$ self-consistently. In our calculations we perform the search for both the self-consistent $\mathcal{G}_0$ and $\mu$ in one shot using a global minimization procedure. As a result the self-consistent solution can be reached in typically less that 20 iterations.

\section{Self-energy and distribution function in the atomic limit}
\label{app_atomic}

In the case $U\gg t$ one can treat the Hamiltonian Eq.~(\ref{H}) using a
perturbative expansion in $t_{ij}/U$. The atomic limit is the zeroth-order term
of this development, and it corresponds to a collection of disconnected sites
with four possible states on each site. This limit is not very useful since
there is no hopping and thus no possible transition below the Hubbard energy
$U$. In order to retain the low-energy dynamics of the problem we adopt an
hybrid approach, where the free dispersion is used in the lattice Green's
function together with the self-energy evaluated in the atomic limit. The atomic
self-energy is obtained by diagonalizing the Hamiltonian Eq.~(\ref{H}) with
$t_{ij}=0$, which leads to the atomic Green's function
	\begin{equation}\label{Gatomic}
		G_{\text{at}}(i\omega_n)=\frac{1-n/2}{i\omega_n+\mu_{\text{at}}}+
		\frac{n/2}{i\omega_n+\mu_{\text{at}}-U},
	\end{equation}
while the non-interacting $G_{0,\text{at}}=1/(i\omega_n+\mu_{\text{at}})$
results by putting $U=0$. From Dyson's equation, $\Sigma=G_0^{-1}-G^{-1}$, we
deduce the atomic self-energy displayed in Eq.~(\ref{atomic-self}). Here
$\mu_{\text{at}}$ is the chemical potential in the \emph{true} atomic
limit---\textit{i.e.} the limit where the lattice Green's function takes the
form (\ref{Gatomic}), and therefore the electron density is given by
$n=(2-n)f(-\mu_{\text{at}})+nf(U-\mu_{\text{at}})$ with $f$ the Fermi function.
We can invert this relation and express $\mu_{\text{at}}$ explicitly in terms of
the electron density as
	\begin{equation*}
		\mu_{\text{at}}=-\frac{1}{\beta}\log\left[{\textstyle\frac{1}{n}-1+
		\sqrt{\left(\frac{1}{n}-1\right)^2+
		e^{-\beta U}\left(\frac{2}{n}-1\right)}}\right].
	\end{equation*}
Using the atomic self-energy Eq.~(\ref{atomic-self}) as an approximation to the
exact self-energy in Eq.~(\ref{n_k}), we evaluate analytically the lattice
distribution function $\langle n_{\vec{k}}\rangle$. Let's first remark that
	\begin{equation*}
		\frac{1}{i\omega_n-\xi_{\vec{k}}-\Sigma_{\text{at}}(i\omega_n)}=
		\frac{A_{\vec{k}}}{i\omega_n-E^+_{\vec{k}}}+
		\frac{1-A_{\vec{k}}}{i\omega_n-E^-_{\vec{k}}}
	\end{equation*}
with
	\begin{eqnarray*}
		E^{\pm}_{\vec{k}}&=&(\xi_{\vec{k}}\pm\Delta_{\vec{k}}+U-\mu_{\text{at}})/2\\
		A_{\vec{k}}&=&\frac{\xi_{\vec{k}}+\Delta_{\vec{k}}+U-\mu_{\text{at}}}
			{4\Delta_{\vec{k}}}\times\\ \nonumber
			&&\frac{(\xi_{\vec{k}}+\Delta_{\vec{k}}-U+\mu_{\text{at}})
			(\mu_{\text{at}}-U+nU/2)+n\mu_{\text{at}}U}{\xi_{\vec{k}}
			(\mu_{\text{at}}-U+nU/2)+n\mu_{\text{at}}U/2}\\
		\Delta_{\vec{k}}&=&\sqrt{(\xi_{\vec{k}}+U+\mu_{\text{at}})^2+2(n-2)
			(\xi_{\vec{k}}+\mu_{\text{at}})U}.
	\end{eqnarray*}
As a result the Matsubara sum in Eq.~(\ref{n_k}) is easily performed to yield
	\begin{equation}
		\langle n_{\vec{k}}\rangle_{\text{at}}=A_{\vec{k}}f(E^+_{\vec{k}})+
		(1-A_{\vec{k}})f(E^-_{\vec{k}}).
	\end{equation}

Within this approximation it is also straightforward to perform the infinite $U$
limit. Taking into account that both $\mu$ and $\mu_{\text{at}}$ are either of
order $t$ (if $n<1$) or of order $U$ (if $n>1$) we find that $A_{\vec{k}}$
approaches $n/2$ as $U$ increases toward $+\infty$. Likewise, if $n<1$ we have
$E^+_{\vec{k}}\sim U$ and $E^-_{\vec{k}}\sim t$ while if $n>1$ we have
$E^+_{\vec{k}}\sim t$ and $E^-_{\vec{k}}\sim-U$. Hence we find
	\begin{equation*}
		\langle n_{\vec{k}}\rangle_{\text{at}}^{U=\infty}=
		\begin{cases}
		\left(1-\frac{n}{2}\right)f\big[\left(1-\frac{n}{2}\right)\xi_{\vec{k}}-
		\frac{n}{2}\mu_{\text{at}}\big]&n<1\\[1em]
		\frac{n}{2}f\big[\frac{n}{2}\tilde{\xi}_{\vec{k}}-
		\left(1-\frac{n}{2}\right)\tilde{\mu}_{\text{at}}\big]+1-\frac{n}{2}&n>1
		\end{cases}
	\end{equation*}
where we have introduced
$\tilde{\mu}_{\text{at}}\equiv\mu_{\text{at}}-U=-\frac{1}{\beta}\log[(1-n/2)/(n-
1)]$ and $\tilde{\xi}_{\vec{k}}\equiv\varepsilon_{\vec{k}}-\tilde{\mu}$ with
$\tilde{\mu}=\mu-U$. For the purpose of evaluating the high-temperature behavior
of the Hall coefficient at $U=\infty$, we finally expand the distribution
function in powers of $\beta$ following the procedure described in
Sec.~\ref{section:RH0_highT}:
	\begin{equation*}
		\langle n_{\vec{k}}\rangle_{\text{at}}^{U=\infty}=
		\begin{cases}
		\frac{n}{2}-n(1-n)\varepsilon_{\vec{k}}\frac{\beta}{2}
			+\mathcal{O}(\beta^2)&n<1\\[1em]
		\frac{n}{2}-(n-2)(1-n)\varepsilon_{\vec{k}}\frac{\beta}{2}
			+\mathcal{O}(\beta^2)&n>1
		\end{cases}
	\end{equation*}
Comparing with Eq.~(\ref{expand}), which is valid at $U=0$, we see that the only
difference between the high-temperature behaviors of $\RH$ at $U=0$ and
$U=\infty$ is the $n$-dependent slope, and we easily deduce that
	\begin{equation*}
		\RH^{U=\infty}(T\gg t)=
		\begin{cases}
			\frac{T/t}{e}\frac{1}{n(1-n)}\frac{a^2\sqrt{3}}{4}\frac{3}{2+(t'/t)^2}
			&n<1\\[1em]
			\frac{T/t}{e}\frac{1}{(n-2)(1-n)}\frac{a^2\sqrt{3}}{4}\frac{3}{2+(t'/t)^2}
			&n>1
		\end{cases}
	\end{equation*}
By introducing $\delta=|n-1|$ which measures the doping with respect to half-filling,
these two cases can be recast in one single expression shown in Eq.~(\ref{RHinf-T}).

\end{document}